\documentclass[zpreprint,zbstdefault,british]{zeus_Kpaper}

\usepackage{zeusdet_units}

\newcommand{\CL}[1]{$#1\%$~C.L.}

\titlehead{
  ZEUS-Note 2016-001 \\
  June 21, 2016
\vspace{4cm}
}

\zeustitle{Simplified QCD fit method\\ for BSM analysis of HERA data}

\zeusauthor{Oleksii Turkot$^{a}\!\!$,
        Katarzyna Wichmann$^{a}$ and
        Aleksander Filip \.Zarnecki$^{b}$\\
        \small
        $^a$DESY ~ and ~
        $^b$Faculty of Physics, University of Warsaw\\
       }

\date{}

\begin{document}

\maketitle
\thispagestyle{empty}

\abstract{
The high-precision HERA data can be used as an input to a QCD analysis
within the DGLAP formalism to obtain the detailed description of the
proton structure in terms of the parton distribution functions (PDFs).
However, when searching for Beyond Standard Model (BSM) contributions in the
data one should take into account the possibility that the PDF set may
already have been biased by partially or totally absorbing previously
unrecognised new physics contributions. 
The ZEUS Collaboration has proposed a new approach to the BSM
analysis of the inclusive $ep$ data based on the
simultaneous QCD fits of parton distribution functions together with
contributions of new physics processes.
Unfortunately, limit setting procedure in the frequentist
approach is very time consuming in this method, as
full QCD analysis has to be repeated for numerous data replicas.
We describe a simplified approach, based on the Taylor
expansion of the cross section predictions in terms of PDF
parameters, which allowed us to reduce the  calculation time for the BSM
limits by almost two orders of magnitude.
}


\section{Introduction}
\label{sec-intro}

The H1 and ZEUS collaborations measured inclusive 
$e^{\pm}p$ scattering cross sections at HERA from
1994 to 2000 (HERA I) and from 2002 to 2007 (HERA II), 
collecting together a total integrated luminosity of about 1\,fb$^{-1}$.
All inclusive data were recently combined \cite{h1zeus_inc} to create
one consistent set of neutral current (NC) and charged current (CC)
cross-section measurements for $e^{\pm}p$ scattering with unpolarised beams.
The inclusive cross sections were used as input to a QCD analysis
within the DGLAP formalism, resulting in a PDF set
denoted as \mbox{HERAPDF2.0}.

The ZEUS collaboration has recently used the HERA combined data 
to set limits on possible deviations from the Standard Model due to a finite 
radius of the quarks \cite{zeus_rq_paper}.
To take into account the possibility that the new physics contributions
can affect PDF determination, resulting in the bias of the QCD fit results,
the limit-setting procedure was based on a simultaneous QCD fit of PDF 
parameters and the quark radius.
The \CL{95} limits on the effective quark-radius squared, $R_{q}^{2}$,
were 
\begin{eqnarray*}
  -(0.47\cdot 10^{-16} \cm)^2 \; < \;
     R_q^{2} & < & (0.43\cdot 10^{-16} \cm)^2 \; .
\end{eqnarray*}
Taking into account the possible influence of quark radii on the PDF 
parameters turned out to be important - for fixed PDFs the obtained
limits would be too strong by about 10\%.

These limits on the effective quark-radius squared were derived
in a frequentist approach \cite{Cousins:1994yw} using the technique of
replicas.  
The replicas are sets of cross-section values that are
generated by varying all cross sections randomly according to their
known statistical and systematic uncertainties.
For each value of the true quark-radius squared, $R_q^{\rm  2\;True}$,
considered in the limit setting procedure, about 5000
replicas were generated and used as an input to a QCD fit with the PDF
parameters and the quark radius squared treated as free parameters.
With a single QCD fit to the full HERA data set taking on average about
1.5 hour of CPU time, 200~000 fits performed for setting the final
limits in the quark radius analysis required over 30 years of CPU time.
Even when using a high performance computing cluster, processing time
is a limiting factor for possible extensions of the analysis
to other models.
 

\section{Standard QCD+BSM fit}
\label{sec-fit}

As described in the $R_q$ paper \cite{zeus_rq_paper},
the PDFs of the proton are described
at a starting scale of $1.9$ GeV$^2$ in terms of $N_{par} = 14$ parameters.  
These parameters, denoted $p_k$ in the following (or $\boldsymbol{p}$
for the set of parameters), together with the possible contribution of
BSM phenomena (quark form factor $R_q^2$ or CI coupling $\eta$) are
fit to the data using a $\chi^2$ method, with the $\chi^2$ formula
given by:
\begin{equation}
 \chi^2 \left(\boldsymbol{p},\boldsymbol{s},\eta \right) = 
 \sum_i
 \frac{\left[m^i
+ \sum_j \gamma^i_j m^i s_j  - {\mu_{0}^i} \right]^2}
{\left( \textstyle \delta^2_{i,{\rm stat}} +
\delta^2_{i,{\rm uncor}} \right) \,  (\mu_{0}^i)^2}
 + \sum_j s^2_j ~~.
\label{eq:qcdfit}
\end{equation} 
Here $\mu_{0}^{i}$ and $m^i$ are the measured cross-section value
and the pQCD+BSM cross-section prediction at the point $i$.
The quantities $\gamma^{i}_j $, $\delta_{i,{\rm stat}} $ and 
$\delta_{i,{\rm uncor}}$ are the relative correlated 
systematic, relative statistical and relative uncorrelated 
systematic uncertainties of the input data, respectively. 
The components $s_j$ of the vector $\boldsymbol{s}$ represent the
correlated systematic shifts of the cross sections (given in units of
$\gamma^{i}_j $), which are fit to the data together with PDF parameter set
$\boldsymbol{p}$ and the CI coupling $\eta$. 
The summations extend over all data points $i = 1, \ldots N_{data}$
and all correlated systematic uncertainties $j=1, \ldots N_{sys}$.

The dependence of the pQCD+BSM cross-section prediction at the point $i$
on the PDF parameters $\boldsymbol{p}$ and the CI coupling $\eta$ can be
written as: 
\begin{equation}
  m^i =  {\cal Q}(x_i, Q^2_i, \boldsymbol{p},\eta)  ,  \label{fullpred}
\end{equation} 
where $x_i$ and $Q^2_i$ are the kinematic variables corresponding to
the point $i$.

\section{Replica generation}
\label{sec-replica}

Equation (\ref{fullpred}) relating model parameters and cross-section
predictions is also used for the replica generation.
For each replica, the generated value of the cross section 
at the point $i$, $\mu^{i}$, is calculated as
\begin{equation}
 \mu^{i}  = 
 \left[ m_{True}^{i} + \sqrt{\delta^2_{i,{\rm stat}} + \delta^2_{i, {\rm uncor}}} \cdot  \mu_{0}^{i} \cdot r_i \right]
\cdot
\left( 1 + \sum_j \gamma^i_j \cdot r_j  \right),
\label{eq:replica}
\end{equation} 
where variables $r_i$ and $r_j$ represent random numbers from a normal
distribution generated for each data point $i$ and for each source of
correlated systematic uncertainty $j$, respectively.
A set of cross-section values $m_{True}^{i}$ is calculated using the
nominal PDF predictions  (based on the set $\boldsymbol{p_{0}}$ of the
PDF parameters fit to the actual data \cite{h1zeus_inc} without taking CI
contribution into account) and the assumed CI coupling value $\eta^{True}$.
It can be written as
\begin{equation}
  m_{True}^i  =    {\cal Q}(x_i, Q^2_i, \boldsymbol{p_{0}}, \eta^{True}) .
\end{equation}
The set of nominal Standard Model predictions can be defined as
\begin{equation}
  m_{0}^i  =    {\cal Q}(x_i, Q^2_i, \boldsymbol{p_{0}}, 0 )  .\label{fullsm}
\end{equation}
In the simplified approach described below, these predictions will be
used as the reference cross section values.

\section{Simplified QCD fit approach}
\label{sec-simpfit}

The proposed approach is based on the assumption that PDF
parameters resulting from the QCD fit fluctuate only within
relatively small uncertainties from replica to
replica.
Therefor, we assume that the dependence of the cross-section
predictions on the PDF parameters can be approximated by a first order
(linear) Taylor expansion, valid for small parameter variations.
For each data point $i$, we define a vector of derivatives:
\begin{eqnarray}
  \theta^i_{0\;k}
  =  \left.  \frac{\partial m_{0}^i}{\partial  p_k}\right|_{\chi^2 = \chi^2_{min}}
  &  =  &
  \left.  \frac{\partial {\cal Q}(x_i, Q^2_i,\boldsymbol{p},0)}{\partial  p_k}
  \right|_{\boldsymbol{p} = \boldsymbol{p_{0}}} ~~, \label{smderiv}
\end{eqnarray} 
where $k = 1 , \ldots N_{par}$. These derivatives can be 
calculated numerically in the linear approximation as:
\begin{equation}
  \theta^i_{0\;k}  =
  \left.  \frac{\partial {\cal Q}(x_i, Q^2_i,\boldsymbol{p},0)}{\partial  p_k}
  \right|_{\boldsymbol{p} = \boldsymbol{p_{0}}}
  \approx  \frac{ {\cal Q}(x_i, Q^2_i,\boldsymbol{p_0^{+k}},0) 
                - {\cal Q}(x_i,
                Q^2_i,\boldsymbol{p_0^{-k}},0)}{\sigma_k} ~.
  \label{derivapp}
\end{equation} 
Here $\sigma_k$ is the uncertainty of the  fitted PDF parameter $p_0^k$
($k$-th parameter of the vector $\boldsymbol{p_0}$) and the two
parameter vectors $\boldsymbol{p_0^{+k}}$ and $\boldsymbol{p_0^{-k}}$
describe parameter sets resulting from changing parameter  $p_0^k$ by
$\pm \frac{1}{2}\sigma_k$:
\begin{eqnarray}
  \boldsymbol{p_0^{+k}} & = & \left( p_0^1, \ldots ,  p_0^k \!+\!
                   \frac{\sigma_k}{2}, \ldots , p_0^{N_{par}} \right) ,\\
  \boldsymbol{p_0^{-k}} & = & \left( p_0^1, \ldots ,  p_0^k \!-\!
                   \frac{\sigma_k}{2}, \ldots , p_0^{N_{par}} \right) .
\end{eqnarray} 
The simplified formula for the model predictions has a form
\begin{eqnarray}
  \tilde{\cal Q}(x_i, Q^2_i, \boldsymbol{p}, 0) 
  &  = &  m_0^i  + \sum_k   \theta^i_{0 \; k} \cdot   \Delta p^k  ,
                                                           \label{simpsm}
\end{eqnarray}
where $\Delta p^k$ is the shift of the PDF parameter
$p^k$ with respect to the nominal fit result,
$\Delta p^k =  p^k -  p_{0}^{k} $.
By substituting exact formula (\ref{fullpred}) by the approximate
formula (\ref{simpsm}) we can significantly speed-up calculation of
the model predictions $m^i$ in the PDF fitting procedure.

\begin{figure}[tbp]
\begin{center}
  \includegraphics[width=0.45\textwidth]{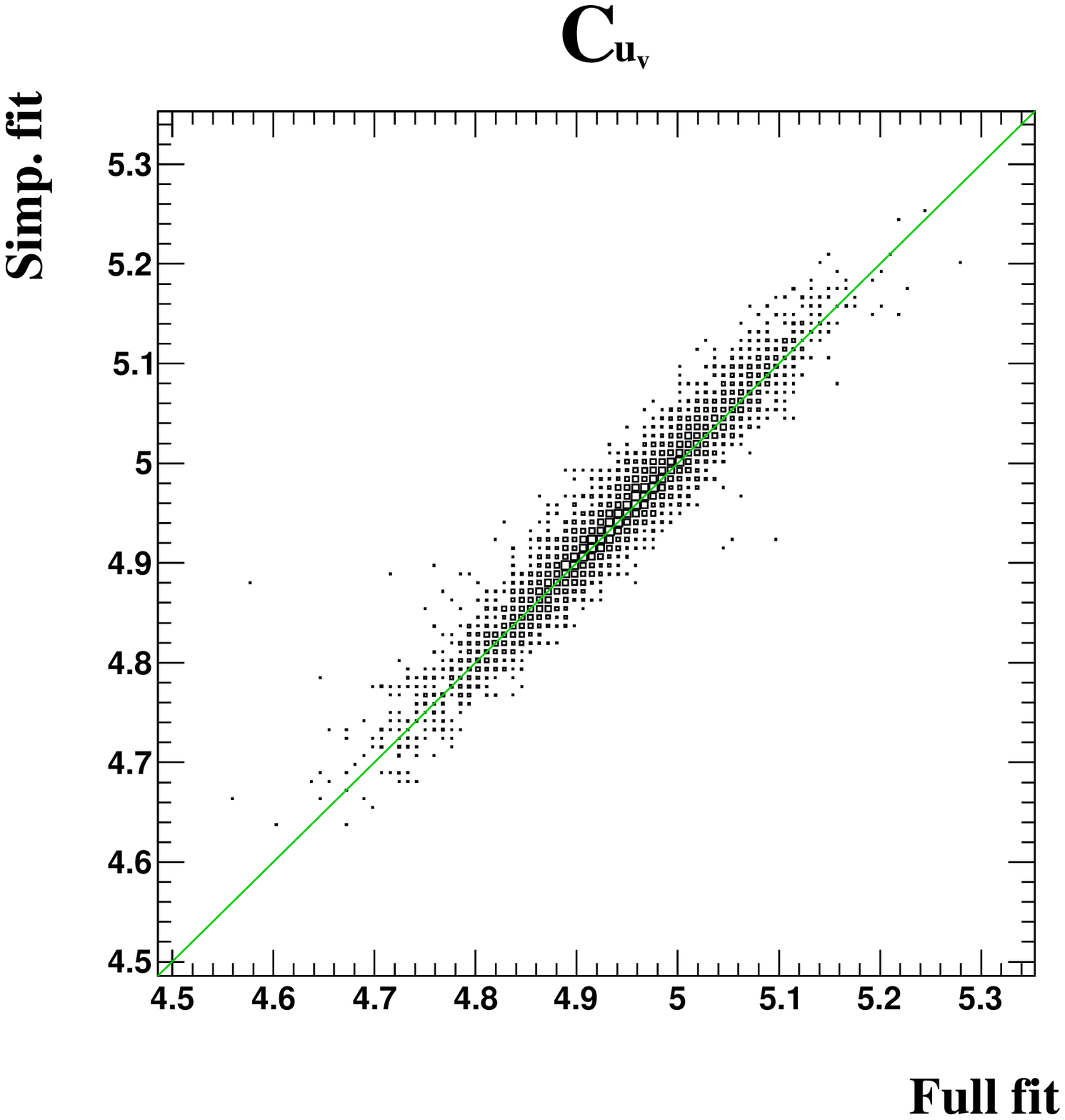}
  \includegraphics[width=0.45\textwidth]{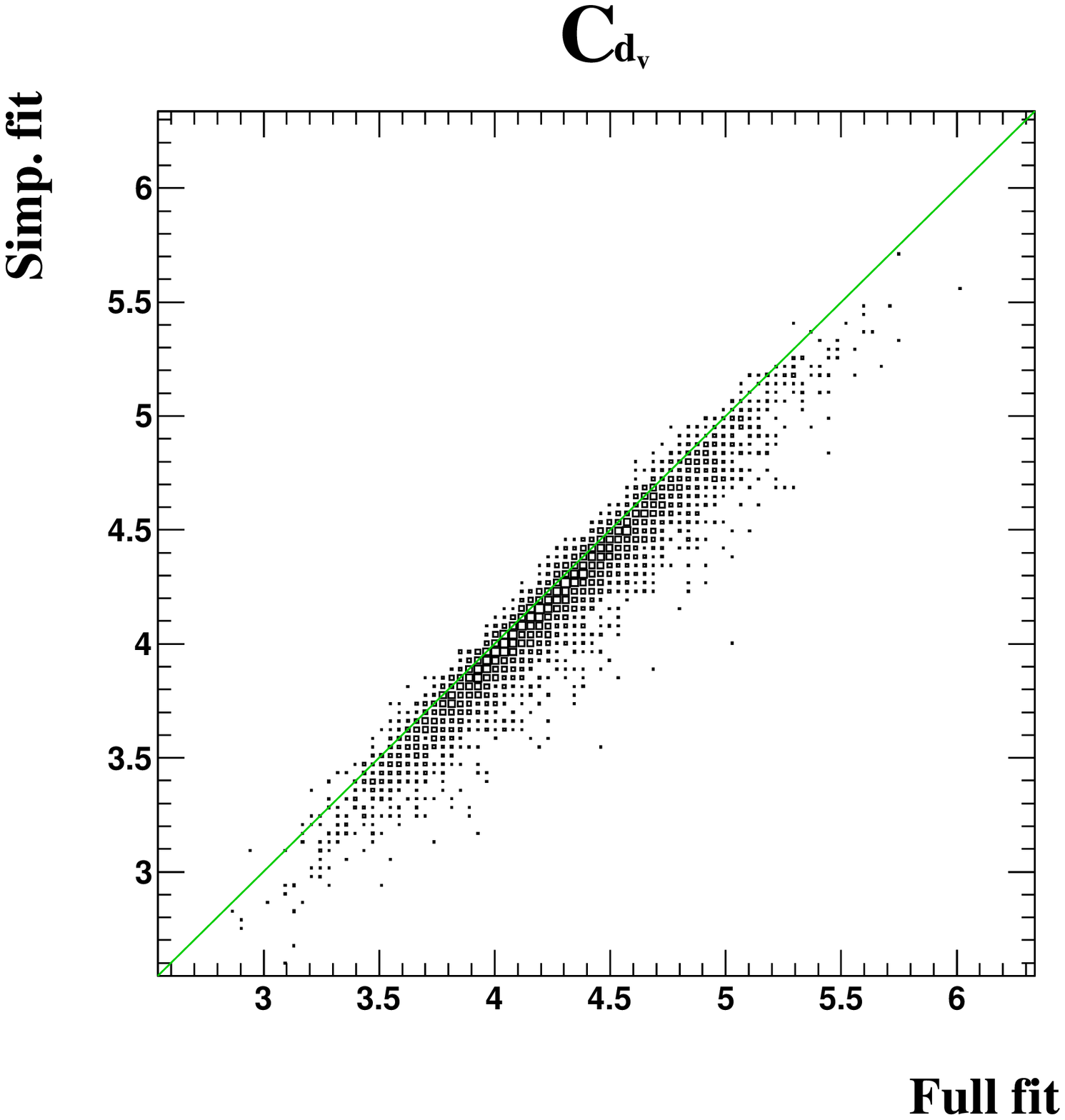}
\end{center}
\begin{center}
  \includegraphics[width=0.45\textwidth]{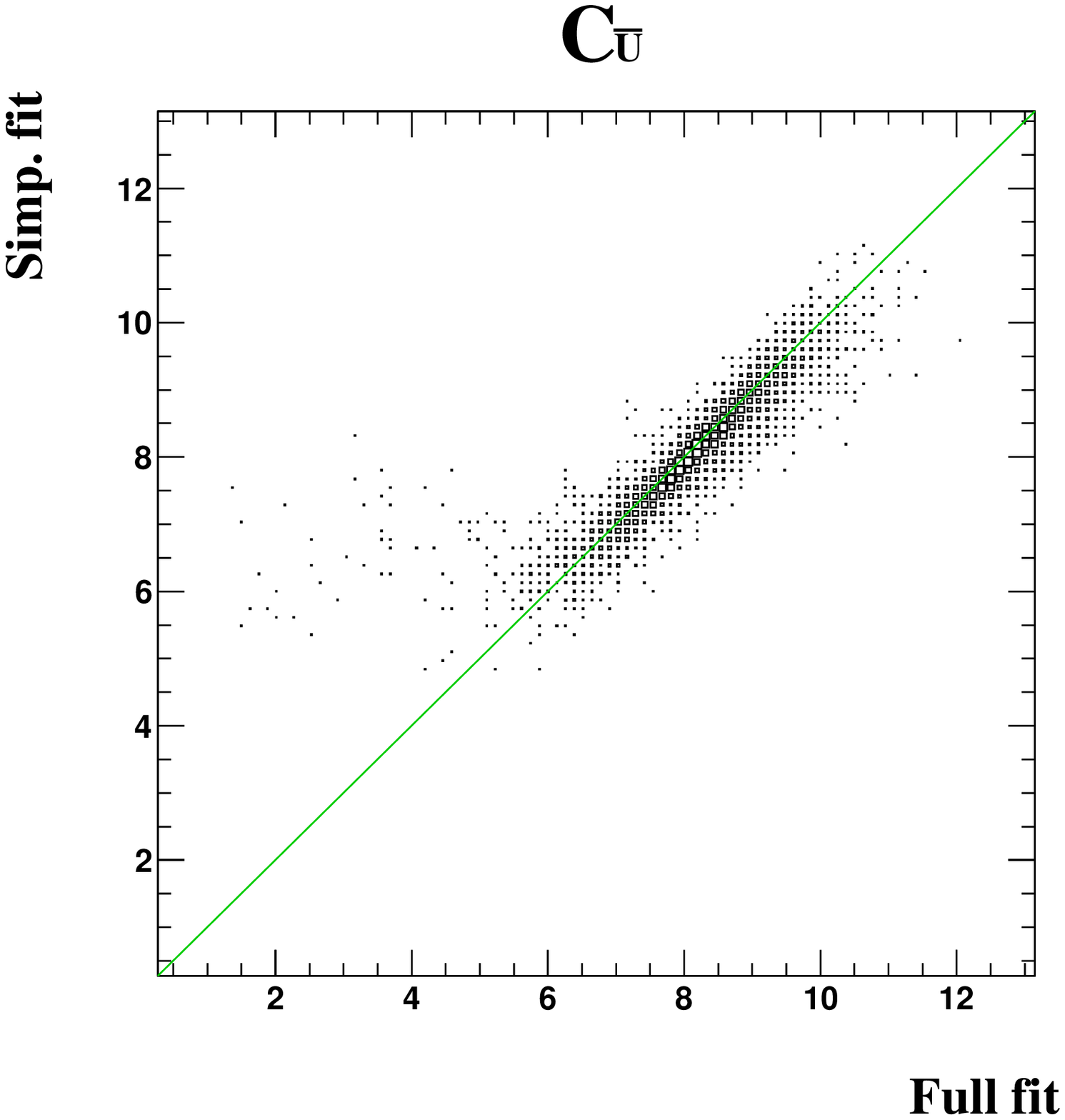}
  \includegraphics[width=0.45\textwidth]{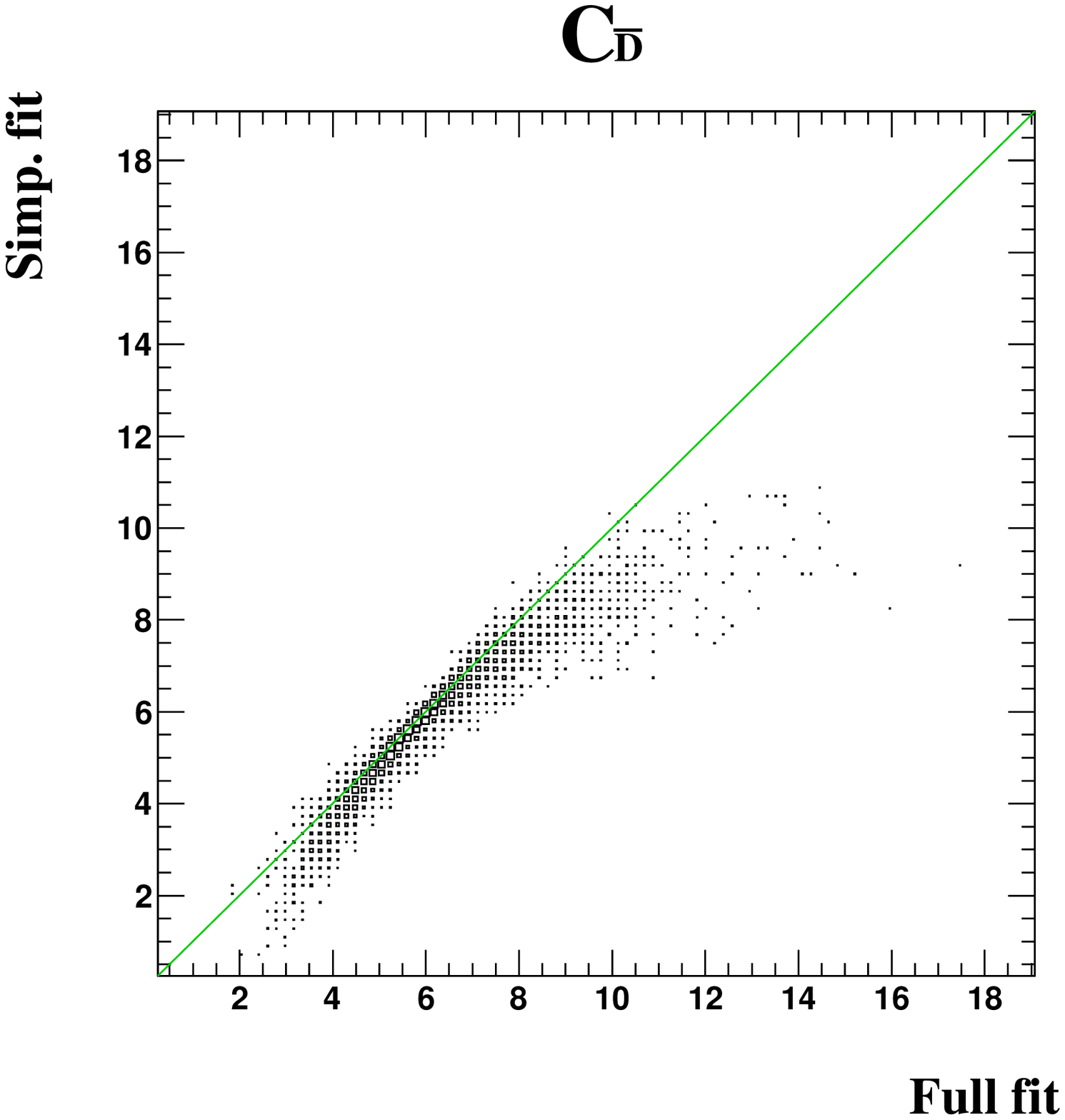}
\end{center}
\caption{Comparison of the chosen PDF parameter values from
  the full QCD fit and from the simplified fit on the large set of the
  Standard Model replicas. Parameters describing high-$x$ behaviour of
  valence and see quark distributions are shown, as indicated
  in the plot labels.
}
\label{smparcorr}
\end{figure}

\begin{figure}[tbp]
\begin{center}
  \includegraphics[width=0.45\textwidth]{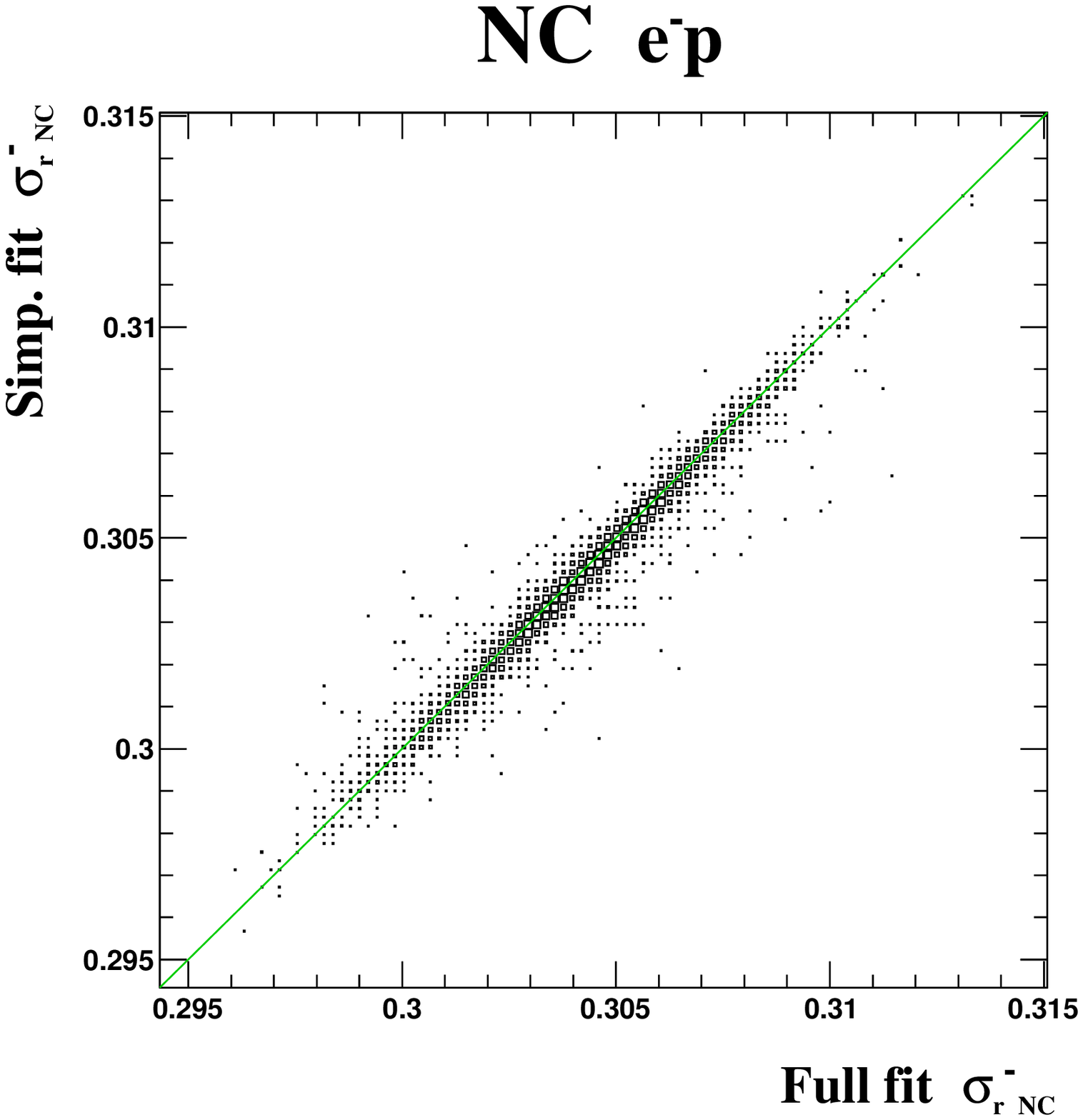}
  \includegraphics[width=0.45\textwidth]{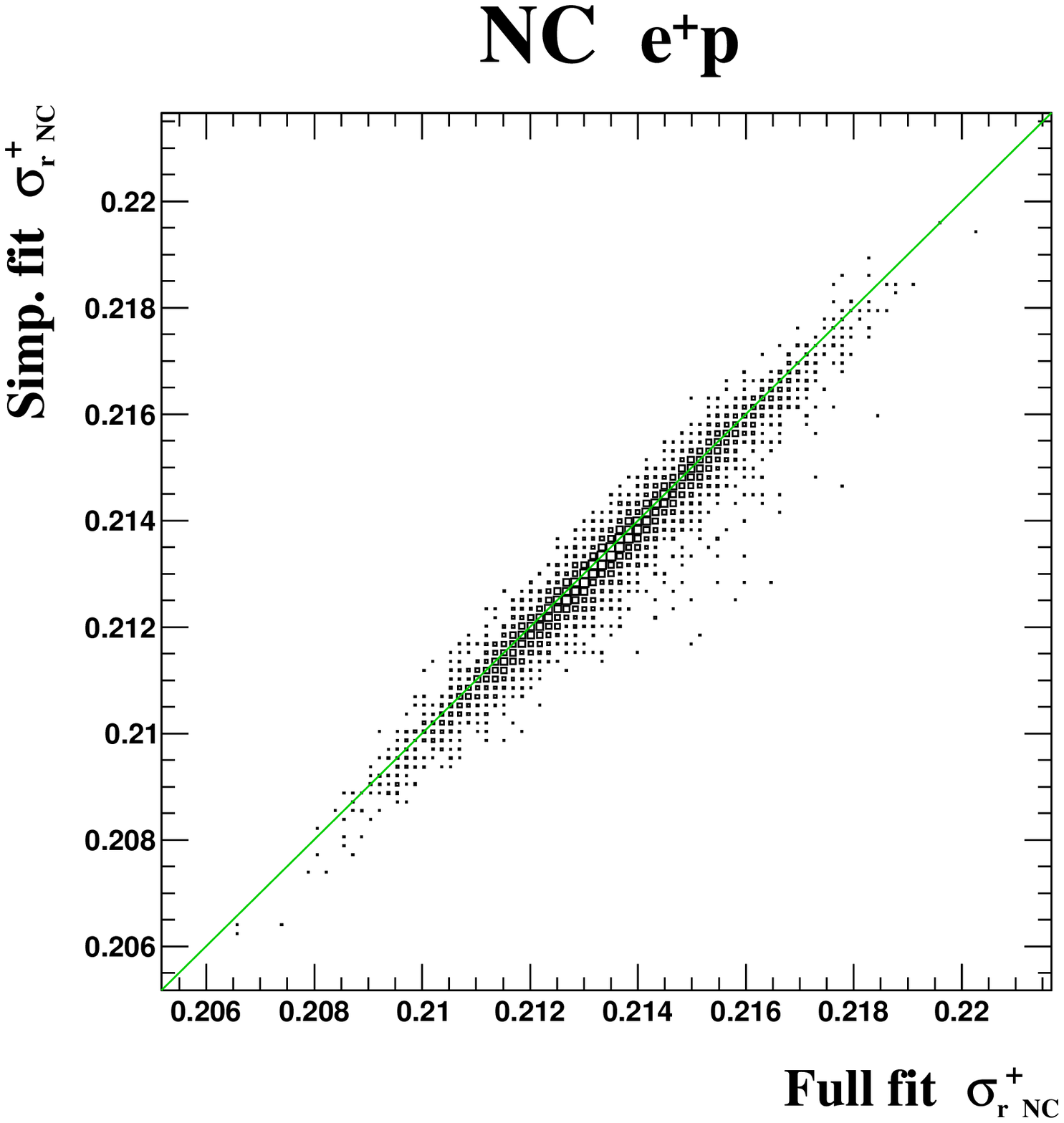}
\end{center}
\begin{center}
  \includegraphics[width=0.45\textwidth]{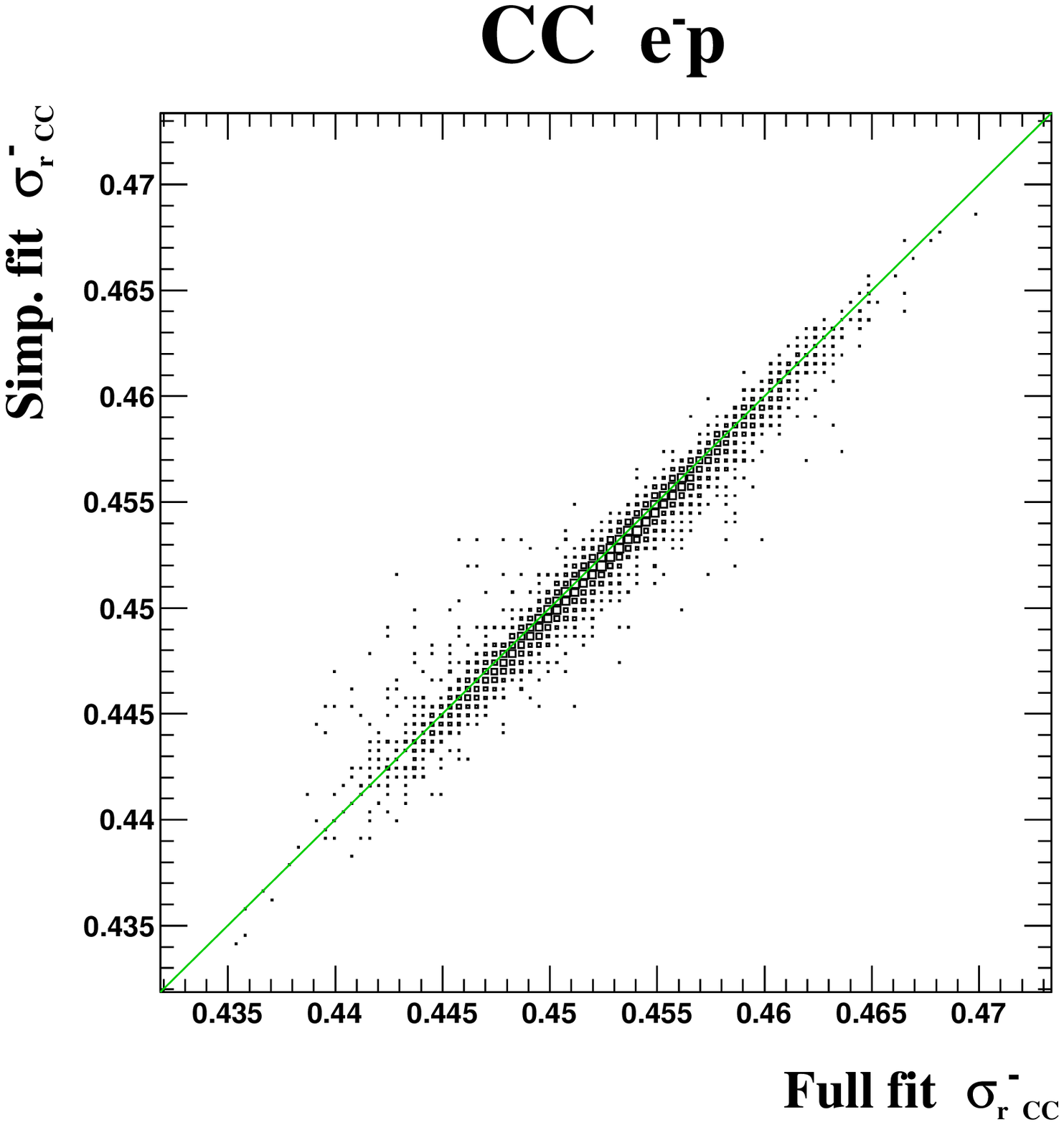}
  \includegraphics[width=0.45\textwidth]{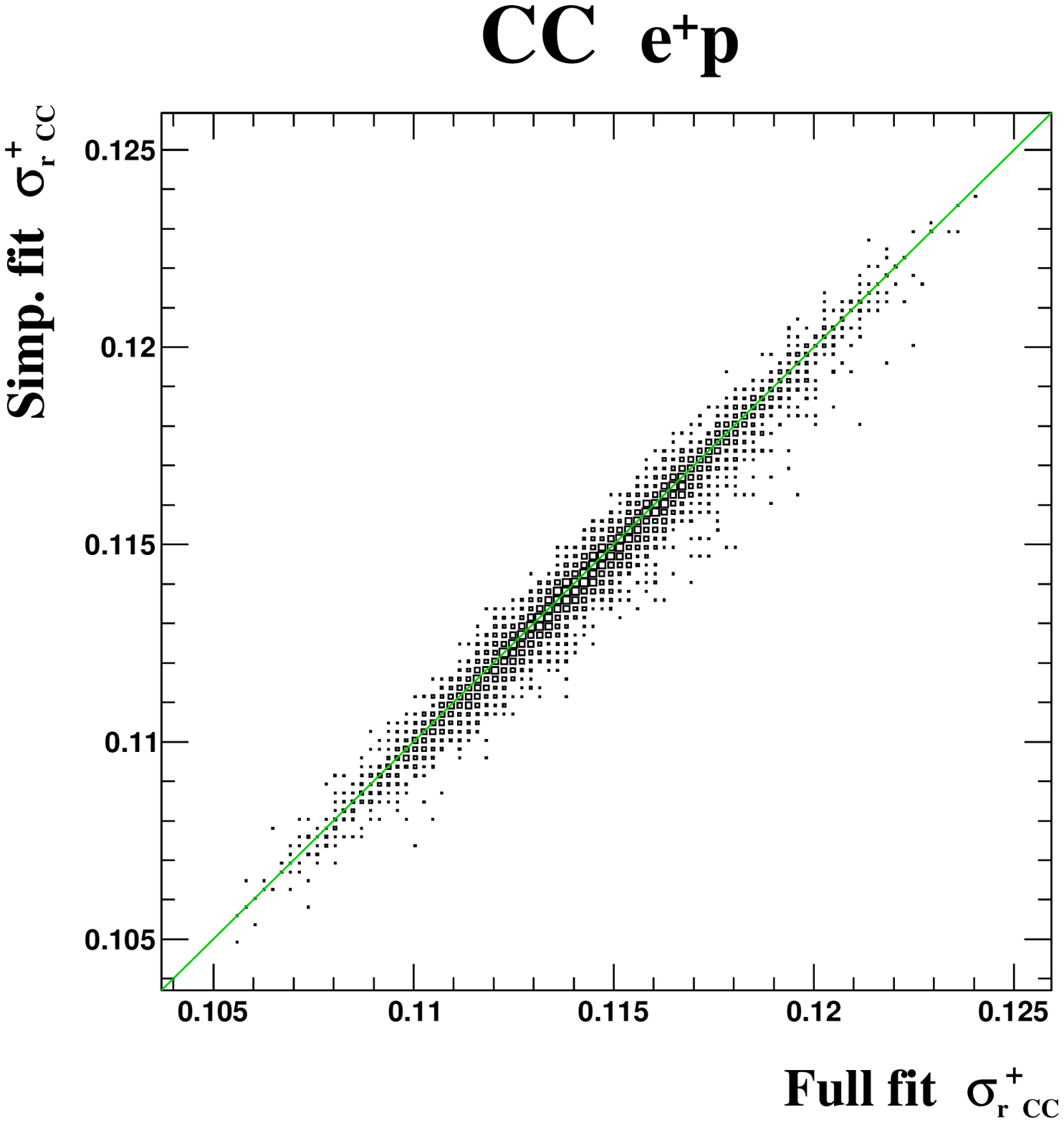}
\end{center}
\caption{Comparison of the reduced cross-section predictions from
  the full QCD fit and from the simplified fit on the large set of the
  Standard Model replicas. Cross sections for NC and CC  $e^{\pm}p$ DIS
  at $x=0.25$ and $Q^2 = 8000$ GeV$^{\: 2}$ are considered, as indicated
  in the plot labels.
}
\label{smfitcorr}
\end{figure}

\begin{figure}[tbp]
\begin{center}
  \includegraphics[width=0.45\textwidth]{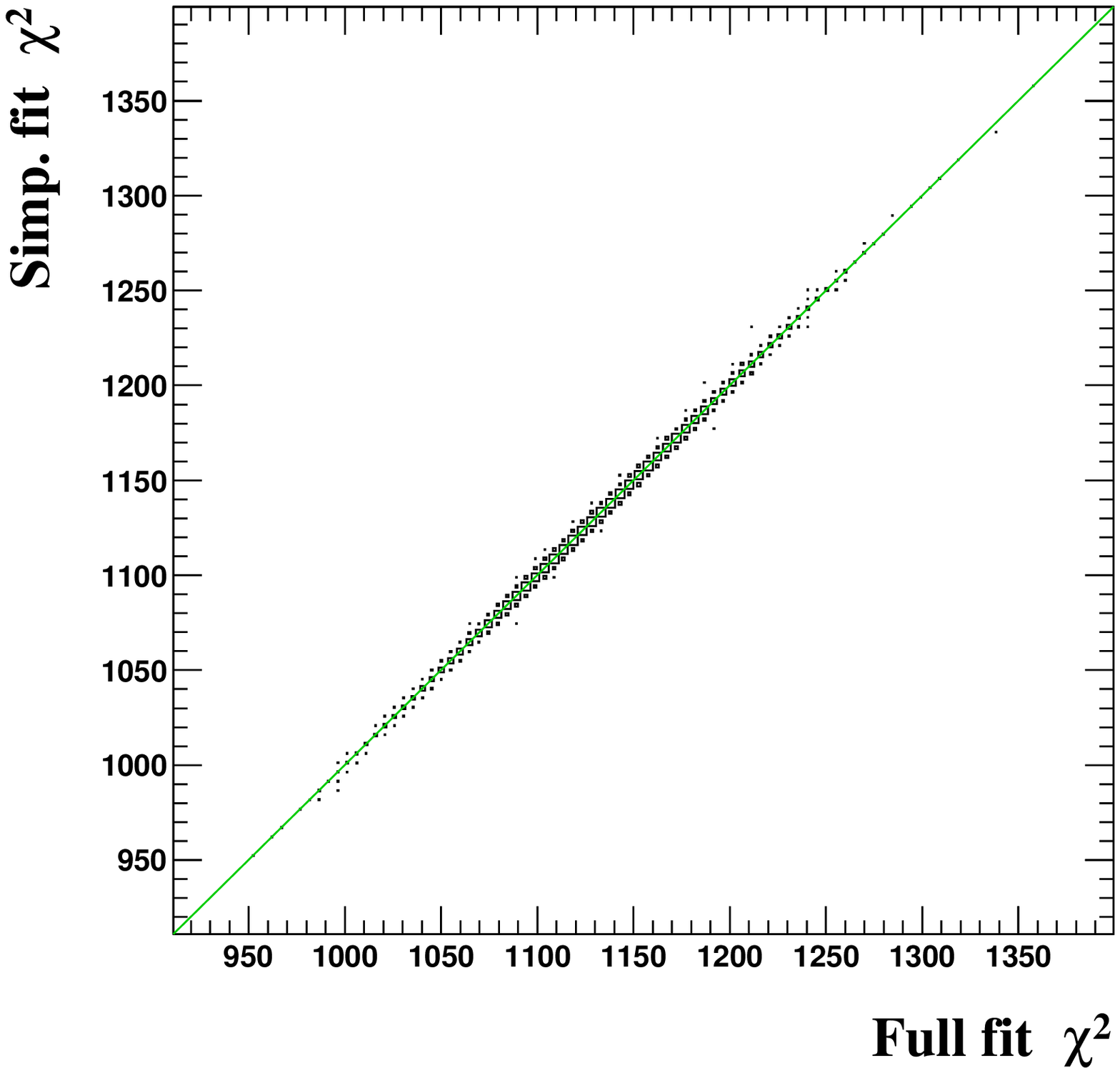}
  \includegraphics[width=0.45\textwidth]{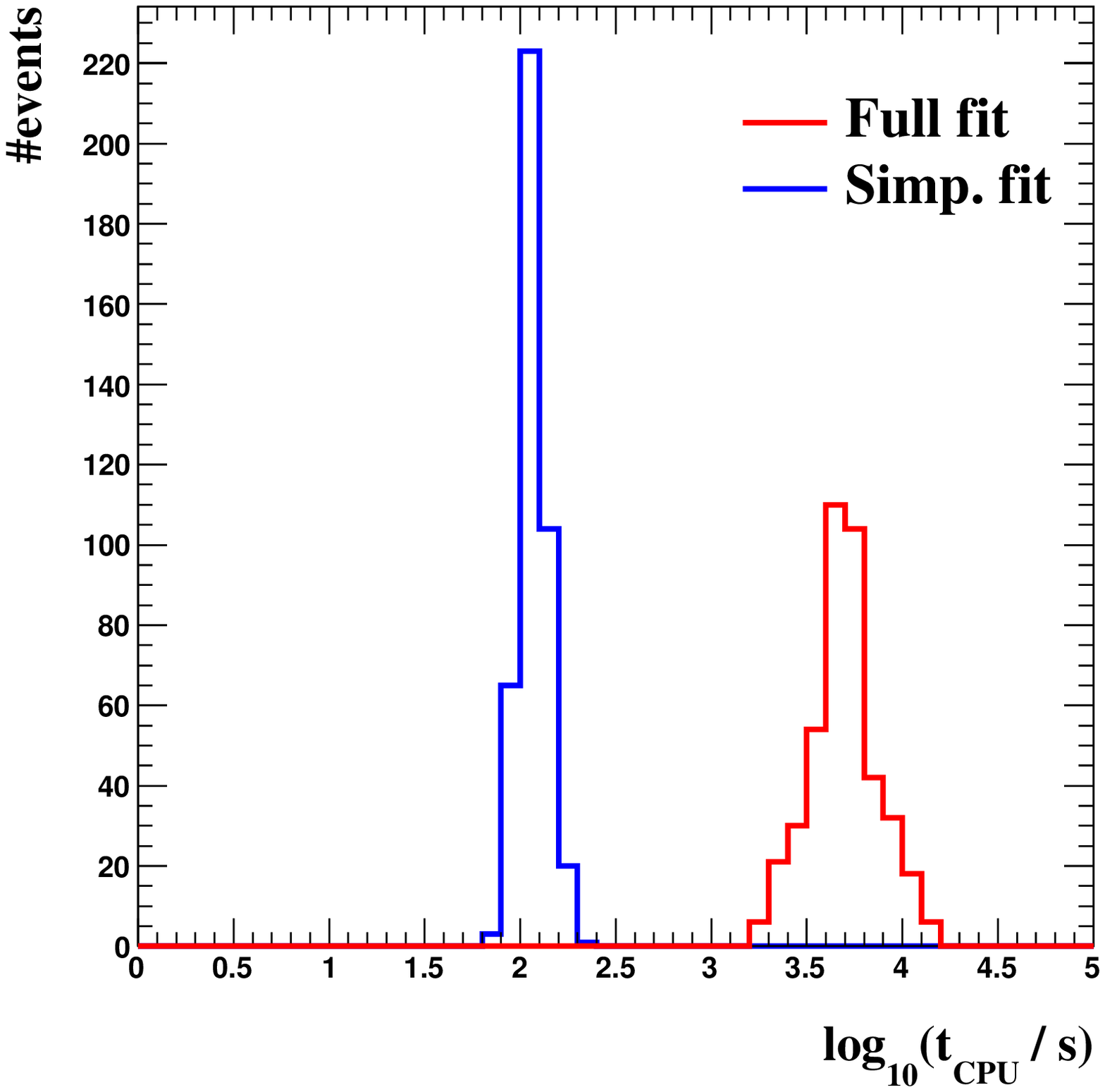}
\end{center}
\caption{Comparison of the full QCD fit performance with the simplified fit
  procedure for  the large set of the  Standard Model replicas,
  for the $\chi^2$ values resulting from the fit (left) and for 
  the CPU time required (right; note the logarithmic scale).
}
\label{cputime}
\end{figure}

The proposed procedure was tested by comparing results of the full QCD
fit and the simplified fit on a large sample of Standard Model replicas
(generated without BSM contribution, i.e. with $\eta^{True}$ set to 0).
Possible CI contribution was also not considered in the fit ($\eta$
parameter fixed to 0).
Parameter values resulting from the full QCD fit and from
the simplified fit on the large set of the Standard Model replicas
are compared in Fig.~\ref{smparcorr}.
Parameters $C_{u_{v}}$, $C_{d_{v}}$, $C_{\bar{U}}$ and $C_{\bar{D}}$
describing high-$x$ behaviour of valence $u$, valence $d$, see $u$ and
see $d$ quarks respectively, are considered.
Distributions of the fitted parameter values agree in general,
but there are also visible differences between the two methods and
a systematic bias for $C_{d_{v}}$ and $C_{\bar{D}}$ parameters.
However, when comparing the reduced cross-section values calculated
from the fitted PDFs, as illustrated in Fig.~\ref{smfitcorr},
the two approaches agree very well.
The simplified fit reproduces results of the full fit with percent
level accuracy and no systematic bias.
The agreement of the replica data with the predictions of DGLAP
evolution equations, as indicated by the $\chi^2$ value of the fit,
is well reproduced
while the processing time is reduced by a factor
of almost 50, as shown in  Fig.~\ref{cputime}.

\section{Simplified QCD+BSM fit}
\label{sec-simpci}

The procedure described above can be easily extended to different CI
scenarios.
Exact description of the pQCD+BSM cross-section predictions as a
function of the coupling parameter $\eta$ can still be preserved.
This is because the dependence of the model predictions on the
coupling $\eta$ is restricted to linear and quadratic terms only.
For each data point $i$, two additional cross-section values
(in addition to the reference value $m^i_{0}$ defined by formula
\ref{fullsm}) can be defined:
\begin{eqnarray}
   m_{+}^i  & = &    {\cal Q}(x_i, Q^2_i, \boldsymbol{p_{0}}, + \Delta \eta ), \\
   m_{-}^i  & = &    {\cal Q}(x_i, Q^2_i, \boldsymbol{p_{0}}, - \Delta \eta  ), 
\end{eqnarray}
where $\Delta \eta$ is a fixed (but otherwise arbitrary) step value
(eg. $\Delta \eta =$ 1 TeV$^{-2}$). These values can be then used to
calculate the cross section terms linear and quadratic in CI coupling:
\begin{eqnarray}
   m_{1}^i  & = &    \frac{m_{+}^i - m_{-}^i}{2\; \Delta \eta } ,\\
   m_{2}^i  & = &    \frac{m_{+}^i + m_{-}^i - 2 \; m_{0}^i}{2\; (\Delta \eta)^2 } .
\end{eqnarray}
The cross section prediction can be then written as
\begin{equation}
  m^{i} =  {\cal Q}(x_i, Q^2_i, \boldsymbol{p_{0}}, \eta )  =
        m_{0}^i  +   m_{1}^i \cdot \eta +   m_{2}^i \cdot \eta^2  . 
\end{equation} 
For each data point $i$, in addition to the three reference
cross-section values ($m_{0}^i$, $m_{+}^i$,$m_{-}^i$) and the vector
of derivatives $\theta^i_{0\;k}$ (see equation \ref{smderiv}),
one needs to calculate two vectors of derivatives, corresponding to $m_{+}^i$ and
$m_{-}^i$ values ($k = 1 , \ldots N_{par}$):
\begin{eqnarray}
  \theta^i_{+\;k}
  =  \left.  \frac{\partial m_{+}^i}{\partial  p_k}\right|_{\chi^2 = \chi^2_{min}}
  &  =  &
  \left.  \frac{\partial {\cal Q}(x_i, Q^2_i,\boldsymbol{p},+\Delta \eta)}{\partial  p_k}
  \right|_{\boldsymbol{p} = \boldsymbol{p_{0}}} ~, \\[5mm]
  \theta^i_{-\;k}
  =  \left.  \frac{\partial m_{-}^i}{\partial  p_k}\right|_{\chi^2 = \chi^2_{min}}
  &  =  &
  \left.  \frac{\partial {\cal Q}(x_i, Q^2_i,\boldsymbol{p},-\Delta \eta)}{\partial  p_k}
  \right|_{\boldsymbol{p} = \boldsymbol{p_{0}}} ~.
\end{eqnarray} 
These derivatives can also be calculated numerically, based on the linear
approximation, see formula (\ref{derivapp}) above.

To summarize, $3 + 3 N_{par}$ reference values have to be stored
for each data point $i$ (three cross section values and three
derivative values for each PDF parameter). These values are calculated
using the full cross section formula (\ref{fullpred}) and the PDF
parameters fit to the {\bf nominal data}. We can then introduce a simplified
description of the CI model predictions:
\begin{eqnarray}
  \tilde{\cal Q}(x_i, Q^2_i, \boldsymbol{p}, \eta ) 
  \;  = \;  m_0^i  + \sum_k   \theta^i_{0 \; k} \Delta p^k 
 & + &  \left( m_1^i  + \sum_{k'}   \theta^i_{1 \; k'} \Delta p^{k'}
 \right) \eta  \nonumber \\
 & + &  \left( m_2^i  + \sum_{k''}   \theta^i_{2 \; k''} \Delta  p^{k''} \right) \eta^2 ,
 \label{cisimperd}
\end{eqnarray}
where $\Delta p^k =  p^k -  p_{0}^{k} $ and $\theta^i_{1 \; k}$,
$\theta^i_{2\; k}$ are combinations of calculated derivatives,
corresponding to cross-section terms linear and quadratic in the coupling:
\begin{eqnarray}
  \theta_{1\;k}^i  & = &
  \frac{\theta_{+\;k}^i - \theta_{-\;k}^i}{2\; \Delta \eta } \\[5mm]
  \theta_{2\;k}^i  & = &
  \frac{\theta_{+\;k}^i + \theta_{-\;k}^i - 2 \; \theta_{0\;k}^i}{2\; (\Delta \eta)^2 } 
\end{eqnarray}
The simplified cross-section function
$\tilde{\cal Q}(x_i, Q^2_i,\boldsymbol{p}, \eta )$
defined by the formula (\ref{cisimperd}) can then replace
the full cross-section calculation (including QCD evolution of
PDFs) given by ${\cal Q}(x_i, Q^2_i,\boldsymbol{p}, \eta )$ of formula (\ref{fullpred})
in the QCD+CI fit procedure for replicas generated for any $\eta^{True}$,
assuming the deviations from nominal Standard Model predictions are small.

\begin{figure}[htbp]
\begin{center}
  \includegraphics[width=0.45\textwidth]{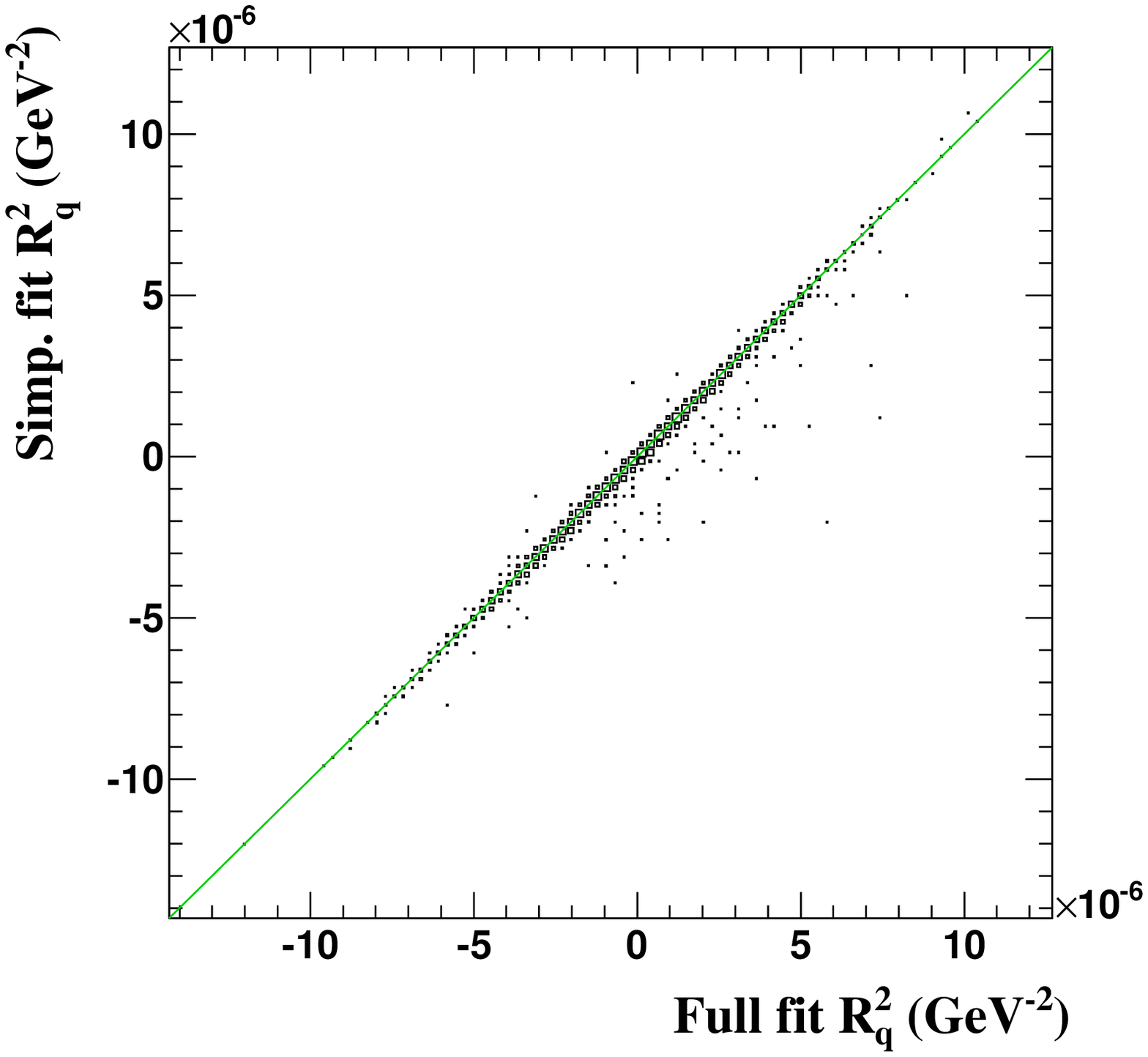}
  \includegraphics[width=0.45\textwidth]{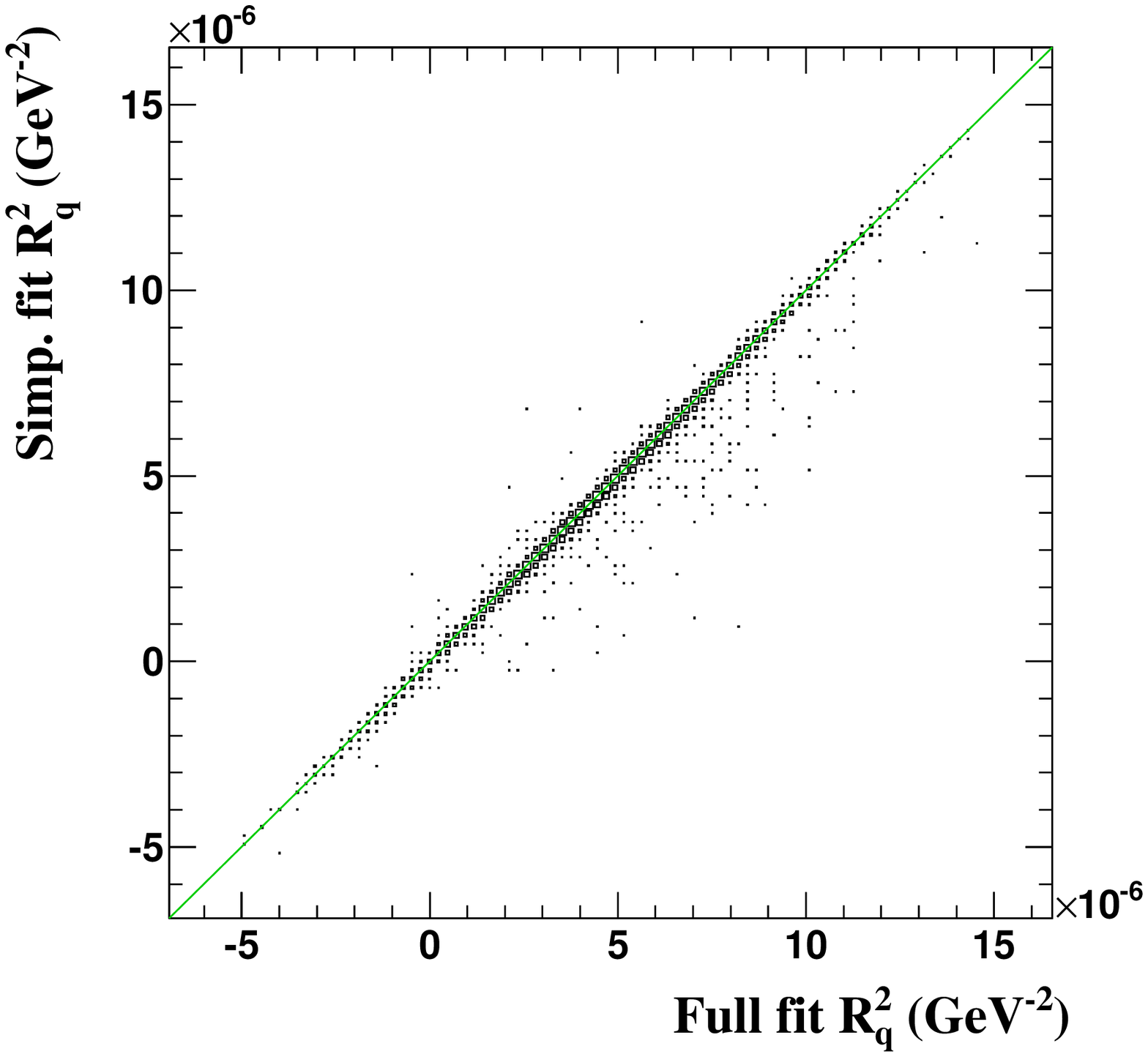}
\end{center}
\caption{Comparison of the quark radius squared, $R_q^{2}$,
  resulting from the full QCD+$R_q$ fit and from the simplified fit
  to the same replica. Results are shown for the set of the Standard
  Model replicas (left) and for the replicas generated with the assumed
  $R_q^{2}$ corresponding to the limit set in the ZEUS
  analysis \protect\cite{zeus_rq_paper} (right).
\label{rqfitcorr}
}
\end{figure}

The presented approach was tested for the quark form-factor model. Shown in
Fig.~\ref{rqfitcorr} is the correlation between the $R_q^2$ value
obtained from the full QCD+$R_q$ fit and those obtained, for the same
replicas, using the simplified approach. The replica sets were generated for the
Standard Model ($R_q \equiv 0$) and for the quark form-factor model with
the quark radius corresponding to the ZEUS limit of
$R_q = 0.43 \cdot 10^{-16}$~cm \cite{zeus_rq_paper}.

\begin{figure}[htbp]
\begin{center}
  \includegraphics[width=0.45\textwidth]{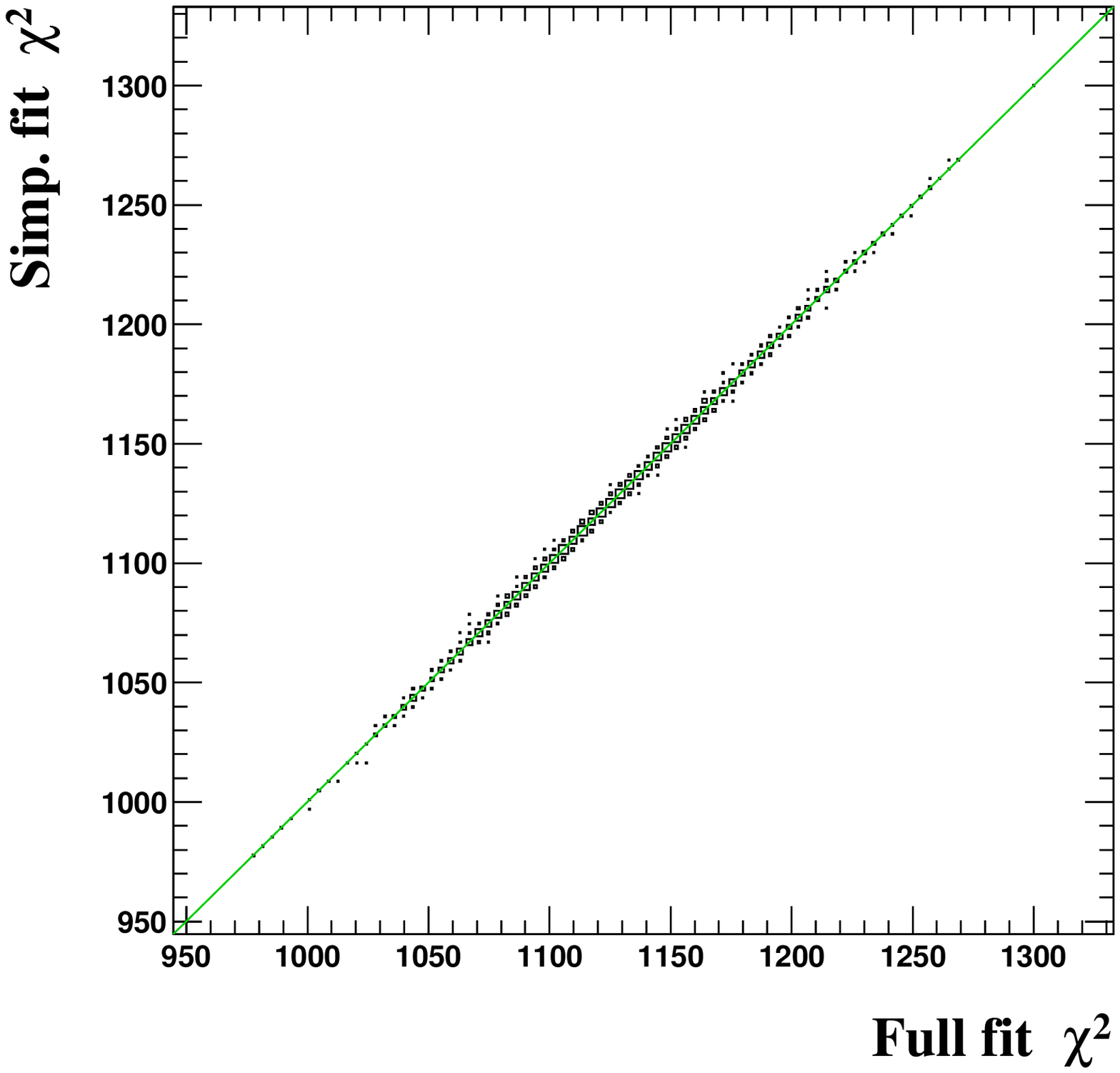}
  \includegraphics[width=0.45\textwidth]{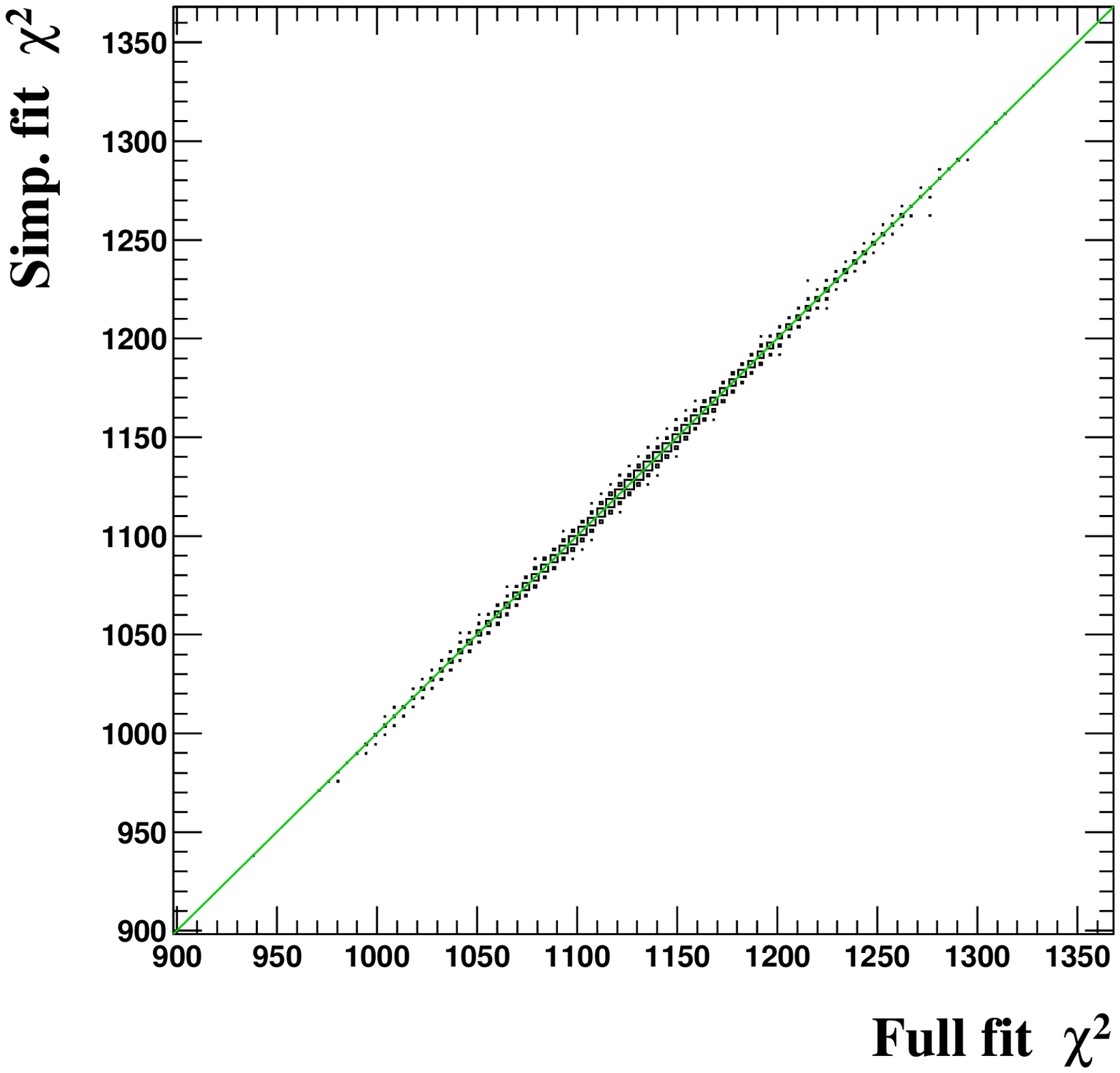}
\end{center}
\caption{Comparison of the  $\chi^2$ values
  resulting from the full QCD+$R_q$ fit and from the simplified fit
  to the same replica. Results are shown for the set of the Standard
  Model replicas (left) and for the replicas generated with the assumed
  $R_q^{2}$ value corresponding to the limit set in the ZEUS
  analysis \protect\cite{zeus_rq_paper} (right).
\label{chi2corr}
}
\end{figure}

\begin{figure}[htbp]
\begin{center}
  \includegraphics[width=0.7\textwidth]{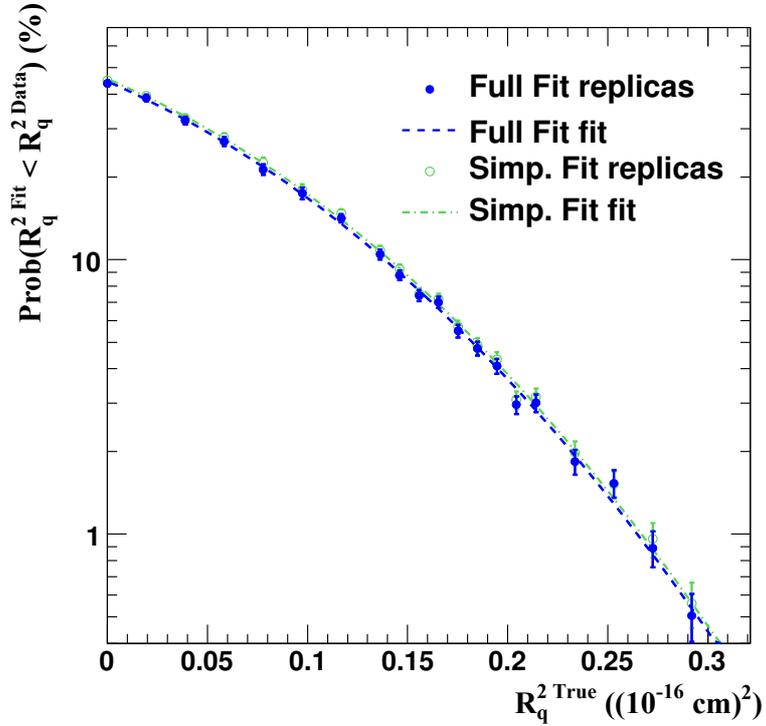}
\end{center}
\caption{
  Results of the limit setting procedure in the frequentist approach,
  based on the QCD fits to multiple data replicas.
The probability of obtaining $R_{q}^{2 \; Fit}$ values smaller 
than that obtained for the actual data, $R_q^{2\; Data}$, is shown
as a function of the assumed value for the quark-radius squared, $R_{q}^{2\;True}$. 
The solid blue circles correspond to the published ZEUS results 
\protect\cite{zeus_rq_paper} obtained with the full QCD+$R_q$ fit
to the replica sets generated for different values of $R_{q}^{2\;True}$,
while the open green circles show the results based on the simplified
fit described in this paper.
The dashed lines represent the cumulative Gaussian distributions fitted
to the replica points.
\label{rqlimit}
}
\end{figure}

As for the cross-section predictions, $R_{q}^{2}$ values fitted with
the simplified method agree almost perfectly with the full QCD+$R_q$
fit results.
Only for a small fraction of replicas some differences are visible,
which are much smaller than the width of the $R_q^2$
distribution.
The quality of the fit, as described by the resulting $\chi^2$ value, is also
very similar for both fit methods, as illustrated in Fig.~\ref{chi2corr}.
When the simplified method is used for the limit setting procedure,
the probability distribution and the resulting limit on the quark
radius squared also agrees very well with the results of
\cite{zeus_rq_paper}, see Fig.~\ref{rqlimit}.

\section{Conclusions}
\label{sec-conclusions}

The simplified procedure for fitting PDF parameters
and BSM couplings to the HERA inclusive data has been developed.
The procedure reproduces the results of the full QCD fit  very well and
allows to shorten the computation time by a factor of 50.
This opens the possibility to extend the quark form-factor analysis \cite{zeus_rq_paper}
of the HERA inclusive data  \cite{h1zeus_inc} to other CI-like scenarios.

\end{document}